\documentclass[twocolumn, aps,prl,showpacs,showkeys,superscriptaddress]{revtex4}
\usepackage[dvips]{graphicx}
\usepackage{indentfirst}
\usepackage{bm}
\usepackage{enumerate}
\usepackage{amsmath}
\usepackage{amssymb}
\usepackage{txfonts}
\usepackage{graphicx}
\usepackage{dcolumn}

\begin{document}

\oddsidemargin=-10mm

\title{Long-distance quantum communication with ``polarization" maximally entangled states}
\author{Fang-Yu Hong}
\affiliation{Department of Physics, Center for Optoelectronics Materials and Devices, Zhejiang Sci-Tech University, Xiasha College Park, Hangzhou, Zhejiang 310018, CHINA}
\author{Shi-Jie Xiong}
\affiliation{National Laboratory of Solid State Microstructures and
Department of Physics, Nanjing University, Nanjing 210093, China}
\author{W.H. Tang}
\affiliation{Department of Physics, Center for Optoelectronics Materials and Devices, Zhejiang Sci-Tech University, Xiasha College Park, Hangzhou, Zhejiang 310018, CHINA}
\date{\today}
\begin{abstract}
 We propose a scheme for long-distance quantum communication where
the elementary  entanglement is generated through two-photon interference  and  quantum swapping is performed through  one-photon interference.  Local ``polarization'' maximally entangled states of atomic ensembles  are generated
by absorbing a single photon from on-demand single-photon sources.  This scheme is robust against  phase fluctuations in the quantum channels, moreover speeds up long-distance high-fidelity entanglement generation rate.
\end{abstract}


\maketitle

Entanglement plays a fundamental role in quantum information science
\cite{pzol} because it is a crucial requisite for quantum metrology
\cite{vgsl}, quantum computation \cite{jcpz,ldhk}, and quantum
communication \cite{jcpz,hbwd}. Quantum communication opens a way
for completely secure transmission of keys with the Ekert protocol
\cite{aeke} and exact transfer of quantum states by quantum
teleportation \cite {chbe}. Because of losses and other noises in
quantum channels, the communication fidelity falls exponentially
with the channel length. In principle, this problem can be
circumvented by applying quantum repeaters \cite{hbwd,lcjt, llsy,
llnm, wmrm}, of which the basic principle is to separate the full
distance into shorter elementary links and to entangle the links with
quantum swaps \cite{chbe,zzhe}. A protocol of special importance for
long-distance quantum communication with collective excitations in
atomic ensembles has been proposed in a seminal paper of Duan {\it
et al.} \cite {dlcz}. After that considerable efforts have been
devoted along this line \cite{crfp,cldc,kchd,nscs,sras,zccs, zyyc}.

In Duan-Lukin-Cirac-Zoller (DLCZ) protocol, entanglement in the elementary links is created by
detecting a single photon from one of two ensembles. The probability
$p$ of generating one excitation in two ensembles is related to the
fidelity of the entanglement, leading to the the condition $p\ll 1$
to guaranty an acceptable quality of the entanglement. But when
$p\rightarrow0$, some experimental  imperfections such as stray
light scattering and detector dark counts will contaminate the
entangled state increasingly \cite{zyyc}, and subsequent processes
including quantum swap and quantum communication become more
challenging for finite coherent time of quantum memory \cite{kchd}.
To solve this problem, protocols based on single photon source
\cite{kchd, nscs} and photon pair source \cite{cshr} were suggested.
However, for the scheme proposed in Ref. \cite{kchd} the ``vacuum"
coefficient $c_0$ \cite{dlcz} of the state of the elementary link is
near 1, which causes the probability $p_i\, (i=1,2,\cdots,n)$ of
successful quantum swap to be very low and thus the capability of
the scheme in increasing quantum communication rate to be weak,
where $n$ is the nesting level of swap. For the schemes suggested in
Refs. \cite{cshr, nscs}, the same problem exists owing to the fact
that the efficiency of storage of a single photon in a quantum
memory is far from ideal. Furthermore, all schemes based on
measuring a single-photon via single-photon detectors suffer from
the imperfections from the detector dark counts and its incapability
of distinguishing one photon from two photons.

Here we present a protocol for long-distance quantum
communication using linear optics and atomic ensembles. To overcome the low probability $p$ in DLCZ protocol,
we generate the entanglement in every node with on-demand
single photon source.  To solve the problem of the large ``vacuum"
coefficient $c_0$ in Refs. \cite{kchd, nscs, cshr}, the quantum
swapping is performed based on ``polarization" maximally entangled
states \cite{dlcz}. Our scheme can automatically eliminate the
imperfection arising from the incapability of the single-photon
detectors in distinguishing one photon from two photons and can exclude partially the imperfection due to the
detector dark counts, which is the major imperfection on the quality
of the entanglement for the previous schemes \cite{nscs}. With this
scheme the quantum communication rate can be significantly increased
by several orders of magnitude with higher quantum communication fidelity
for a distance 2500 km compared with the DLCZ protocol. To be insensitive to the phase fluctuation in the quantum channel \cite{zccs, zcbz}, our previous propose  for quantum communication \cite{fyhsx} employs  two-photon Hong-Ou-Mandel-type (HOMT) interferences to generate local entanglement, to distribute basic entanglement between distance $L_0$, and to connect  entanglement with quantum swap. Because the phase instability in the local quantum channel is easy to control, this scheme uses single-photon Mach-Zehnder-type interferences to generate local entanglement and to connect entanglement, and uses  two-photon HOMT interferences only to distribute basic entanglement to simplify the physical set-up.

\begin{figure}
\includegraphics[width=0.8\columnwidth]{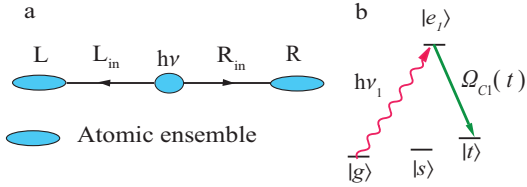}
\caption{\label{fig:1}(Color online) (a) Schematic illustration of
entanglement establishment between two atomic ensembles L and R
through on-demand single-photon sources.  (b) The relevant level
configuration of atoms in the ensembles and the coupling with
pulses.}
\end{figure}

The quantum memory in our scheme can be a cloud of $N_a$ identical
atoms with pertinent level structure shown in Fig. \ref{fig:1} b.
One ground state $|g\rangle$ and two  metastable states $|s\rangle$
and $|t\rangle$ may be provided by, for instance, hyperfine or
Zeeman sublevels of the electronic ground state of alkali-metal
atoms, where long relevant coherent lifetime has been observed
\cite{jhjs, dpaf, twsc}. The atomic ensemble is optically thick
along one direction to enhance the coupling to light \cite{dlcz}.
State $|e_1\rangle$ is an excited state. A single photon emitted with a repetition rate $r$
from an on-demand single-photon source \cite{kchd, sfas} located
halfway between quantum memories $L$ and $R$ in every node is split into an entangled
state of optical modes $L_{in}$ and $R_{in}$ (Fig. \ref{fig:1} a)
described by
\begin{equation}\label{eq4}
  |\psi_{in}(\phi)\rangle= \frac{1}{\sqrt{2}} \left(|0_{L_{in}} \rangle|
  1_{R_{in}}\rangle+{\text e}^{i\phi}|1_{L_{in}}\rangle|0_{R_{in}}\rangle\right)
  \end{equation}
where $\phi$ denotes an unknown difference of the phase shifts in the $L$ and $R$ side channels.
This state then is coherently  mapped onto the state of atomic
ensembles $L$ and $R$:
\begin{equation}\label{eq5}
|\psi(\phi)\rangle_{LR}=  \frac{1}{\sqrt{2}}\left(T_L^\dagger+{\text e}^{i\phi}T_R^\dagger\right)|0_a\rangle_L|0_a\rangle_R
 \end{equation}
by applying techniques such as adiabatic passage based on dynamic
electromagnetically induced transparency \cite{kchd}, where
$T\equiv1/\sqrt{N_a}\sum_{i=1}^{N_a}|g\rangle_i\langle t|$ is the
annihilation  operator for the symmetric collective atomic mode $T$
\cite{dlcz} and $|0_a\rangle\equiv\otimes_i|g\rangle_i$ is the
ensemble ground state.  Considering photon loss, which includes the
optical absorption in the quantum channel and the inefficiency of
the excitation transfer from the optical mode to quantum memory
mode, the state of  ensembles R and L can be described by an
effective maximally entangled (EME) state \cite{dlcz}
 \begin{equation}\label{eq7}
    \rho_{LR}(c_0,\phi)=\frac{1}{c_0+1}\left(c_0|0_a0_a\rangle_{LR}\langle0_a0_a|+
    |\psi(\phi)\rangle_{LR}\langle\psi(\phi)|\right)
 \end{equation}
 where $c_0$ is the vacuum coefficient.

\begin{figure}
\includegraphics[width=0.8\columnwidth]{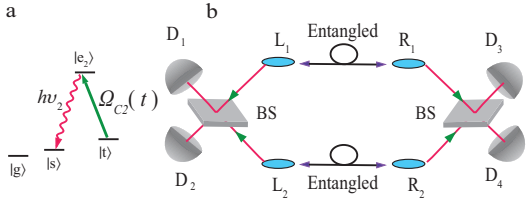}
\caption{\label{fig:2} (Color online)(a) The relevant level
configuration of the atoms in the ensemble and coupling pulses. (b)
Configuration for projecting an effective maximally entangled (EME)
state to  a "polarization" maximally entangled (PME) state.}
\end{figure}

Before proceeding we discuss the conversion of the collective atomic excitation  $T$ into the atomic excitation $S$ given by
$S\equiv1/\sqrt{N_a}\sum_{i=1}^{N_a}|g\rangle_i\langle s|$.
 Consider the atoms have an excited
state $|e_2\rangle$ satisfying the condition that the dipole moments of the atomic transitions $
e\textbf{r}_{1}=e\langle g|\textbf{r}|e_2\rangle=0$, $
e\textbf{r}_{2}=e\langle s|\textbf{r}|e_2\rangle\neq0$, and $
e\textbf{r}_{3}=e\langle t|\textbf{r}|e_2\rangle\neq0$ \cite{mfai}. The transition $|s\rangle\rightarrow|e_2\rangle$ of each of these atoms is coupled to a quantized radiation mode described by an annihilation operator $\hat{a}$  with a coupling constant $g$; the transitions from $|e_2\rangle\rightarrow|t\rangle$ are resonantly driven by a classical control field of Rabi frequency $\Omega_{c2}$ (Fig.\ref{fig:2}). The interaction Hamiltonian of this systems is in the form \cite{mlsy}
\begin{equation}\label{eq8}
 H_{in}=\hbar g \hat{a}\sum_{i=1}^{N} \hat{\sigma}_{e_2s}^i+\hbar \Omega_{c2}(t)\sum_{i=1}^{N} \hat{\sigma}_{e_2t}^i +H.c.
\end{equation}
where $\sigma_{\mu\nu}^i=|\mu\rangle_{ii}\langle\nu|$ is the flip operator of the $i$th atom between states $|\mu\rangle$ and $|\nu\rangle$. This interaction Hamiltonian has the dark state with zero adiabatic eigenvalue \cite{appm, mlsy, mfml},
\begin{equation}\label{eq8}
    |D\rangle=\cos\theta(t) S^\dag|g\rangle|1\rangle-\sin \theta(t)T^\dag |g\rangle|0\rangle,
\end{equation}
where $\tan\theta=g/\Omega_{c2}(t)$ and $|m\rangle$ denotes the radiation state with $m$ photon. Thus with this dark state,
by applying a retrieval pulse of suitable polarization that is
resonant with the atomic transition
$|t\rangle\rightarrow|e_2\rangle$, the atomic excitation $T$ in an
atom ensemble can be converted into the atomic excitation $S$ while a
photon which has polarization and frequency different from the
retrieval pulse is emitted  \cite{dlcz,mlsy,dpaf,mfai,clzd,fyhsx}. Because this conversion process does not involve the collective enhancement, its efficiency  is low.

Now we discuss the generation of local entanglement. Two pairs of ensembles are prepared in the same EME state $
\rho_{L_iR_i}\,(i=1,2)$ at every node with the vacuum coefficient $c_0$ (Fig.\ref{fig:2}).  The
$\phi$ parameters in $ \rho_{L_1R_1}$ and $ \rho_{L_1R_1}$ are equal
assuming that the two EME states are generated through the same
stationary channels. The state of the two pairs of ensembles can be
described with $ \rho_{L_1R_1}\otimes \rho_{L_2R_2}$. By applying
retrieval pulses on resonance with the atomic transition
$|t\rangle\rightarrow|e_2\rangle$, the atomic excitations $T$ are
transformed simultaneously into excitations $S$ while photons are
emitted. After the conversion, the stimulated photons overlap at a
50\%-50\% beam split (BS), and then are recorded by the single-photon
detectors $D_{L_1}$, $D_{L_2}$ ( $D_{R_1}$, $D_{R_2}$) which measures the combined radiation from two samples, $a^\dag_{+L}a_{+L}$ or  $a^\dag_{-L}a_{-L}$ ($a^\dag_{+R}a_{+R}$ or  $a^\dag_{-R}a_{-R}$), with $a_{\pm i}=a_{1i}\pm \text{e}^{i\phi_i}a_{2i},i=L,R $ \cite {dlcz}. In the following discussion, we assume $\phi_L=\phi_R$, which is easy to control for the local transformation \cite{zcbz, ccjl}.  Only the
coincidences of the two-side detectors are recorded, so the protocol
succeeds with a probability $p_r$ only if both of the detectors on
the left and right sides have a click. Under this circumstance, the
vacuum components in the EME states, the state components
$T^\dag_{L_1}T^\dag_{L_2}|vac\rangle$, and
$T^\dag_{R_1}T^\dag_{R_2}|vac\rangle$ have no effect on the
experimental results, where $|vac\rangle$ is the ground state of the
ensemble $|0_a0_a0_a0_a\rangle_{L_1R_1L_2R_2}$. Thus, after the
conversion, the state of system of four ensembles can be written as
the following polarization maximally entangled (PME) state
\begin{equation}\label{eq2}
    |\psi\rangle_{PME}^\pm=(S^\dag_{L1}S^\dag_{R2}\pm S^\dag_{L2}S^\dag_{R1})|vac\rangle/\sqrt{2}.
  \end{equation}
Without loss of generality, we assume that the generated PME is
$|\psi\rangle_{PME}^+$  in the following discussion. The success probability for
entanglement generation at every node is $p_r=\eta_p^2\eta_s^2\eta_{e_1}^2\eta_d^2/2$, where we denote the probability of emitting one photon by the
single-photon source with $\eta_p$, the efficiency for the
atomic ensemble storing a photon by $\eta_s$, the efficiency for the
atomic ensemble emitting a photon during the process $T^\dag|0_a\rangle\rightarrow S^\dag|0_a\rangle$ by $\eta_{e_1}$, and  the single-photon detection efficiency by $\eta_d$. The average waiting time for successful generating a local entanglement state is $T_l=\frac{1}{rp_r}$.

Then we show how to  distribute basic entanglement between neighboring nodes at a distance $L_0$. The atomic ensembles at neighboring nodes $A$ and $B$ are prepared in the state $|\psi\rangle_{PME}^+$, then illuminated  simultaneously by retrieval  laser pulses on resonance of the atomic transition
$|s\rangle\rightarrow|e_3\rangle$, where $|e_3\rangle$ an excited state, the atomic excitations $S$ are
transformed simultaneously into  anti-Stokes photons. We assume the anti-stokes photons are in an orthogonal polarization state $|H\rangle$ from ensemble $A_{R1},B_{L1}$ and $|V\rangle$ from ensemble $A_{R2},B_{L2}$, which represent horizontal and vertical linear polarization, respectively.

\begin{figure}
\includegraphics[width=0.8\columnwidth]{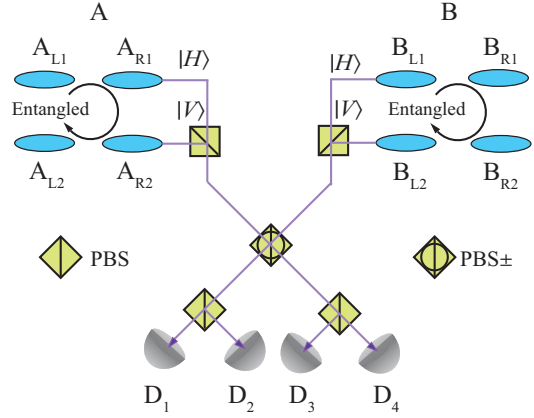}
\caption{\label{fig:3} (Color online) Schematic illustration of basic
entanglement generation with length $L_0$. Up to a local unitary phase shift the coincidence count between single-photon detectors $D_1$ and $D_4$ ($D_1$ and $D_3$) or $D_2$ and $D_3$ ($D_2$ and $D_4$ will project the atomic ensembles at $A_{Li}$ and $B_{Ri} (i=1,2)$ into a PME state in the form of equation \eqref{eq1}. PBS (PBS$_\pm$) transmits  $|H\rangle$ ($|+\rangle$) photons and reflects  $|V\rangle$ ($|-\rangle$) photons, where $|\pm\rangle=\frac{1}{\sqrt{2}}(|H\rangle \pm |V\rangle)$.  }
\end{figure}

  After the conversion, the stokes photons from site $A$ and $B$ at every node are directed to the polarization beam splitter (PBS) and experience two-photon Bell-state measurement (BSM) (shown in Fig.\ref{fig:3}) at the middle point to generate an entanglement between the atomic ensembles $A_{Li}$ and $B_{Ri}$ (i=1,2).
   Only the
coincidences of the two single-photon detectors $D_1$ and  $D_4$ ($D_1$ and  $D_3$) or $D_2$ and  $D_3$ ($D_2$ and  $D_4$) are recorded, so the protocol
is successful only if each of the paired detectors have a click. Under this circumstance, the
vacuum components in the EME states, one-excitation components like $S^\dag_{L_u}|vac\rangle$, and the two-excitation components
$S^\dag_{A_{L1}}S^\dag_{B_{R1}}|vac\rangle$ and
$S^\dag_{A_{L2}}S^\dag_{B_{R2}}|vac\rangle$ have no effect on the
experimental results \cite{kmhw}. A coincidence click between single-photon detectors, for example, $D_1$ and  $D_4$  will project the four atomic ensembles into PME state \cite{zcbz,qzxb,kmhw}
  \begin{equation}\label{eq1}
    |\Psi\rangle^+_{AB}=\frac{1}{\sqrt{2}}(S^\dag_{A_{L1}}S^\dag_{B_{R2}}+S^\dag_{A_{L2}}S^\dag_{B_{R1}})|vac\rangle.
  \end{equation}
    The success probability for
entanglement generation within the attenuation length is $p_b=\eta_{e2}^2\eta_d^2\eta_t^2/2$, where $\eta_{e2}$ denotes the efficiency for the
atomic ensemble emitting a photon during the process  $S^\dag|0_a\rangle\rightarrow |0_a\rangle$ and $\eta_t=\text{exp}[-L_0/(2L_{att})]$ is the fiber transmission
efficiency with the attenuation length $L_{att}$.

\begin{figure}
\includegraphics[width=0.8\columnwidth]{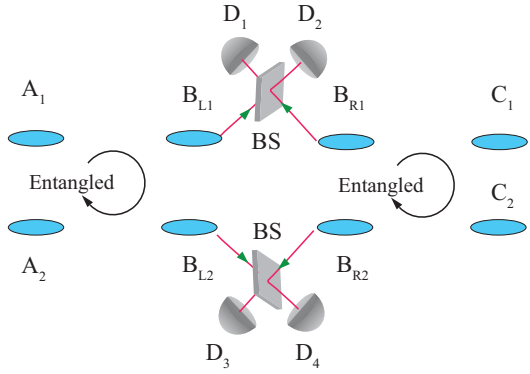}
\caption{\label{fig:4} (Color online) Configuration for entanglement swapping.}
\end{figure}

After successful generation of PME states within the basic link, we
can extend the quantum communication distance through entanglement
swapping with the configuration shown in Fig.\ref{fig:4}. We
have two pairs of ensembles---$A_1$, $A_2$, $B_{L_1}$ and $B_{L_2}$, and
$B_{R_1}$, $B_{R_2}$, $C_1$ and $C_2$, --- located at three sites A, B, and
C. Each pair of ensembles is prepared in the PME state (Eq. \eqref{eq1}). The stored
atomic excitations of four ensembles $B_{L_1}$ and $B_{L_2}$, and
$B_{R_1}$, $B_{R_2}$ are transferred into light at the same time with
near unity efficiency. The stimulated optical excitations interfere
at a 50\%-50\% beam splitter, and then are detected by single-photon
detectors $D_1$, $D_2$, $D_3$, and $D_4$. Only if each pair of
detectors ($D_1$,  $D_2$), and ($D_3$, $D_4$), has a click, the
protocol is successful with a probability $p_1=\eta_{e2}^2\eta_d^2/2$ and a PME state in
the form of equation \eqref{eq1} is established among the ensembles
$A_1$, $A_2$, $C_1$, and $C_2$ with a doubled communication
distance. Otherwise, we need to repeat the previous processes
.

The scheme for entanglement swapping can be applied to arbitrarily
extend the communication distance. For the $i$th ($i=1,2,...,n$)
entanglement swapping, we first prepare simultaneously two pairs of
ensembles in the PME states (Eq. \eqref{eq1}) with the same communication length
$L_{i-1}$, and then make entanglement swapping as shown by Fig.
\ref{fig:4} with a success probability $p_i=\eta_{e2}^2\eta_d^2/2$. After a successful
entanglement swapping, a new PME state is established and  the
communication length is extended to $L_i=2L_{i-1}$. Since the $i$th
entanglement swapping needs to be repeated on average $1/p_i$ times,
the average total time needed to generating a PME state over the
distance $L_n=2^nL_0$ is given by \cite{cshr}
 \begin{equation}\label{eq6}
T_{tot}=\left(\frac{L_0}{c}+\frac{1}{rp_r}\right)\frac{1}{p_b\prod_{i=1}^mp_i}\left(\frac{3}{2}\right)^n
\end{equation}
 with $c$ being the light speed in the optical
fiber.

\begin{figure}
\includegraphics[width=0.8\columnwidth]{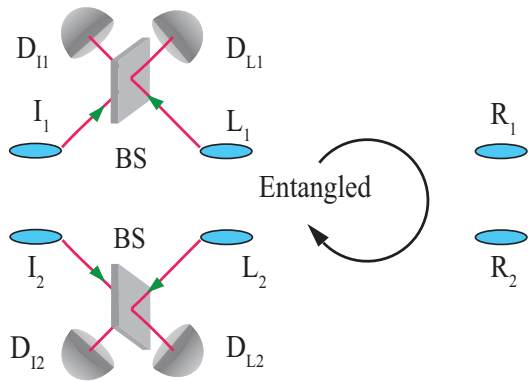}
\caption{\label{fig:5} (Color online) Configuration for
probabilistic quantum teleportation of an unknown atomic
``polarization" state.}
\end{figure}

After a PME state has been generated between two remote sites,
quantum communication protocols, such as cryptography and Bell
inequality detection, can be performed with that PME state like the DLCZ scheme \cite{dlcz}. The established long-distance PME state
can be used to faithfully transfer unknown state through quantum
teleportation with the configuration shown in Fig. \ref{fig:5}. Two
pairs of atomic ensembles $L_1$, $R_1$ and $L_2$, $R_2$ are prepared
in the PME state. The unknown state which is to be transferred is
described by $(\alpha S^\dag_{I_1}+\beta
S^\dag_{I_2})|0_a0_a\rangle_{I_1I_2}$ with unknown coefficient
$\alpha$ and $\beta$, where $S^\dag_{I_1}$ and $S^\dag_{I_1}$ are
the collective atomic operators for the two ensembles $I_1$ and
$I_2$. The collective atomic excitations in the ensembles $I_1$,
$L_1$ and $I_2$, $L_2$  are transferred  into optical excitations
simultaneously. After a 50\%-50\% beam splitter, the optical
excitations are measured by detectors $D_{I_1}$,$D_{L_1}$ and
$D_{I_2}$,$D_{L_2}$. Only if there is one click in
$D_{I_1}$,$D_{L_1}$ and one click in $D_{I_2}$,$D_{L_2}$, the state
transfer is successful, and the unknown state  $(\alpha
S^\dag_{I_1}+\beta S^\dag_{I_2})|0_a0_a\rangle_{R_1R_2}$ appears in
the ensembles $R_1$ and $R_2$ up to a local $\pi$-phase rotation. Unlike the DLCZ protocol, this scheme does not need posterior confirmation of the presence of the excitation to teleportation unknown state.

 Now we evaluate the perform of the scheme numerically.  The conversion efficiency $\eta_{e_1}$ may be low, assuming to be 0.01.  If we assume that $r=50$ MHz, $\eta_p=1$, $\eta_{s}=\eta_{e_2}=0.9$,
$\eta_d=0.9$, $L_n=2500$ km, $L_{att}=22$ km for photons with
wavelength of $1.5 \mu$m \cite{nscs}, $c=2.0\times10^5$ km/s, and
$n=4$, equation (\ref{eq6}) gives the average total time
$T_{tot}=2251$ s, in contrast to  the average total time
$T_{tot}=650000$ s for the DLCZ protocol and $T_{tot}=15300$ s for
single-photon source (SPS) protocol \cite{nscs} with the above
parameters. Thus, compared with the  SPS protocol, this scheme can significantly
reduce the average total time for successful quantum communication.  Note that $e_2$ can be enhanced by putting the atomic ensembles in a low-finesse ring cavity \cite{dlcz} and  one can exploited many kinds of on-demand single-photon sources, such as molecule-based sources with max rate $100$ MHz and quantum-dot-based sources with max rate 1 GHz \cite{blmo}.

 To enhance the conversion efficiency $\eta_{e1}$, we can use a cavity with a quality factor $Q$. According to the the literature \cite{dlcz}, in the free-space limit the signal-to-noise ratio $R_{sn}$ between the coherent interaction rate and the decay rate can be estimated as
 \begin{equation}\label{eq9}
    R_{sn}\sim\frac{4N_a|g_c|^2}{\kappa\gamma_s}\sim3\frac{\rho_nL_a}{k_s^2}\sim d_o,
 \end{equation}
 where $\rho_n$ and  $d_o$ denote the density and  the on-resonance optical depth of the atomic ensemble, respectively, $L_a$ is the length of the pencil-shape atomic ensemble, $k_s=\omega_s/c=2\pi / \lambda _s$ is the wave vector of the cavity mode, and $\kappa$ is the cavity decay rate.  $\kappa$ is relate to the quality factor $Q$ of the cavity $\kappa=\omega_s/Q$ \cite{msmz}. Thus for the case of the cavity with a quality factor $Q$, we have the signal-to-noise ratio
 \begin{equation}\label{eq10}
    R_{sn}\sim\frac{4N_a|g_c|^2}{\kappa\gamma_s}\sim3\frac{\rho_nL_a}{k_s^2}Q\sim3\frac{\rho_nL_a\lambda_s^2}{4\pi^2}Q\sim d_o,
 \end{equation}
which shows that the cavity quality factor $Q$ and the atom number of the ensemble $N$ play a similar role in enhancing the atom-photon interaction.
To estimate the magnitude of the signal-to-noise ratio $R_{sn}$,  we assume $3\rho_nL_a\lambda_s^2/4\pi^2\sim10^{-2}$ for the case of a single atom. Then we have $R_{sn}\sim 10\sim d_o $ for $Q=1000$. According to the research \cite{agaa}, the maximum total efficiency for a single photon storage in an atomic ensemble followed by retrieval can be larger than 0.5 for $d_o=10$. Thus, the conversion efficiency $\eta_{e1}$ larger than 0.01 is feasible if the atomic ensemble is placed in the cavity with a quality factor $Q=1000$.

Now we discuss imperfections in our architecture for quantum
communication.  In the basic entanglement generation, the contamination of
entanglement from processes containing two excitations can be
arbitrarily suppressed with unending advances in single-photon
sources \cite{sfas,blmo}. In the whole process of basic entanglement
generation, connection, and entanglement application, the photon
loss includes contributions from channel absorption, spontaneous
emissions in atomic ensembles, conversion inefficiency of
single-photon into and out of atomic ensembles, and inefficiency of
single-photon detectors. This loss decreases the success probability
but has no effect on the fidelity of the quantum communication
performed. Decoherence from dark counts in the basic entanglement generation and  the entanglement connection can be
excluded, for example, if a dark count occurs on the up side
($D_{1}$ and $D_{2}$) (Fig. \ref{fig:4}), because in this case there
are two clicks in the down side detectors ($D_{2}$ and $D_{4}$), thus
the protocol fails and the previous steps need to be repeated. Considering that the probability  for a detector to give a dark count denoted by $p_d$ smaller than $5\times10^{-6}$ is within the reach of the current techniques \cite{nscs}, we can estimate the fidelity imperfection $\Delta F\equiv1-F$ for the generated long-distance PME states by
\begin{equation}\label{eq3}
    \Delta F=2^{n+2}p_d<3.2\times 10^{-4}
\end{equation}
 for $n=4$.

The imperfection that the detectors cannot distinguish between one
and two photons only reduces the probability of successful
entanglement generation and connection, but has no influence on both
of the fidelity of the PME state generated and the quality of
quantum communication. For instance, if two photons have been
miscounted as one click in detectors $D_{I1}$ and $D_{L1}$ in Fig.\ref{fig:5}, then there is no click in the detectors $D_{I2}$ and
$D_{L2}$, thus the protocol says that the state transfer fails. Like
DLCZ protocol, the  phase shifts arising from the stationary quantum
channels and the small asymmetry of the stationary set-up can be
eliminated spontaneously when we generate the PME state from the EME
state, and thus have no effect on the communication fidelity.
 Because the basic entanglement between distance $L_0$ is generated through two-photon interference, this scheme is robust against the phase fluctuation in the quantum channels \cite{zcbz}.

In conclusion, we  have proposed a robust scheme for long-distance quantum
communication based on ``polarization" maximally entangled state.
Through this scheme, the rate of long-distance quantum communication
may increase  compared with
the  SPS protocol. At the same time,
higher fidelity of long-distance quantum communication can be
expected. Considering the simplicity of the physical set-ups used,
this scheme may opens up the probability of efficient long-distance
quantum communication.

{\it Acknowledgments} This work was supported by the State Key
Programs for Basic Research of China (2005CB623605 and
2006CB921803), by National Foundation of Natural Science in
China Grant Nos. 10474033 and 60676056, and the National Nature Science Foundation of China (Grant No. 50672088 and 60571029).

\end{document}